\documentclass{article}
\usepackage{spconf,amsmath,graphicx}

\usepackage{enumitem}

\usepackage[T1]{fontenc}
\usepackage{amsmath}
\usepackage{graphicx}
\usepackage{caption}

\usepackage{amssymb}
\usepackage{multirow}
\usepackage[table,xcdraw]{xcolor}
\usepackage{graphicx,verbatim}
\usepackage{subcaption}
\usepackage{color}

\setlist{nosep, leftmargin=14pt}

\usepackage{mwe} 

\title{\textit{Gaze2Report}: Radiology Report Generation via Visual-Gaze Prompt Tuning of LLMs}
\name{Aishik Konwer \qquad Moinak Bhattacharya \qquad Prateek Prasanna}
\address{Stony Brook University, Stony Brook, NY, USA}
\begin{document}
%
\maketitle
\begin{abstract}
Existing deep learning methods for radiology report generation enhance diagnostic efficiency but often overlook physician-informed medical priors. This leads to a suboptimal alignment between the structured explanations and disease manifestations. Eye gaze data provides critical insights into a radiologist's visual attention, enhancing the relevance and interpretability of extracted features while aligning with human decision-making processes. However, despite its promising potential, the integration of eye gaze information into AI-driven medical imaging workflows is impeded by challenges such as the complexity of multimodal data fusion and the high cost of gaze acquisition, particularly its absence during inference, limiting its practical applicability in real-world clinical settings. To address these issues, we introduce \textit{Gaze2Report}, a framework which leverages a scanpath prediction module and Graph Neural Network (GNN) to generate joint visual-gaze tokens. Combined with instruction and report tokens, these form a multimodal prompt used to fine-tune LoRA layers of large language models (LLMs) for autoregressive report generation. \textit{Gaze2Report} enhances report quality through eye-gaze-guided visual learning and incorporates on-the-fly scanpath prediction, enabling the model to operate without gaze input during inference. 
\end{abstract}

\section{Introduction}
The rapid growth of medical imaging data has significantly increased the diagnostic workload for radiologists, necessitating fast yet precise reporting to ensure timely clinical decision-making. Automated report generation~\cite{automate1,automate2} is an active area of research involving deep learning systems that generate text descriptions from radiological images to assist physicians in diagnosis and treatment planning. Most approaches follow the standard image captioning architecture~\cite{caption1,caption2}, where a visual encoder extracts image features and passes them to a text decoder to generate reports. These frameworks focus solely on the imaging modality, overlooking the need for additional context beyond a single image for accurate report generation. As a result, they may enhance Natural Language Generation (NLG) metrics but fail to maintain the factual and clinical accuracy. Natural image-captioning approaches are also optimized to describe visual scenes concisely with brief sentences~\cite{caption3}. They fail to capture the contextual depth required in longer, more detailed paragraphs typically found in radiology reports.\\
\begin{figure*}[t]
\centering
\includegraphics[width=.8\textwidth]{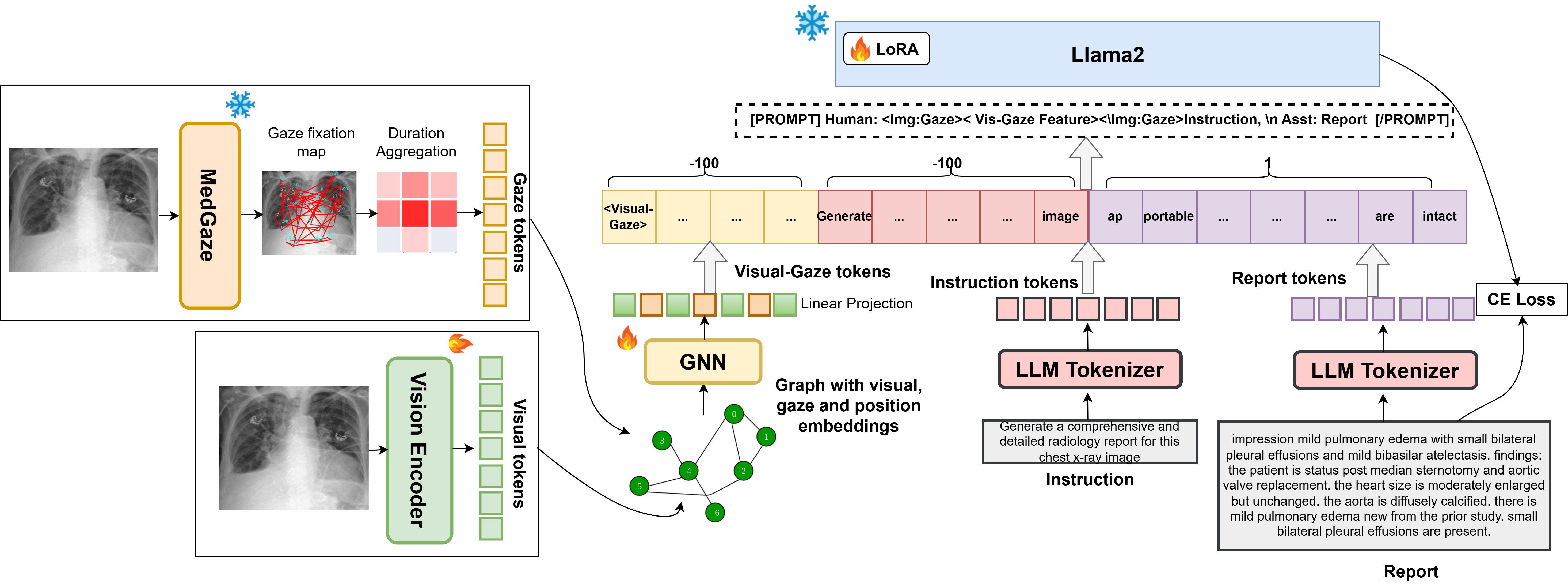}
\caption{Overview of the \textit{Gaze2Report} architecture. -100 means those tokens are not involved in autoregressive training.} \label{framework}
\end{figure*}
To address this, few approaches have been developed to incorporate prior clinical knowledge or attention to clinical abnormalities to improve report accuracy and ensure factual correctness~\cite{camanet}. 
Despite injecting priors and utilizing cross-attention mechanisms to identify abnormalities, these models struggle to effectively integrate and reason over complex multimodal data. This restricts their ability to capture the nuanced relationships between clinically abnormal regions and conclusions in generated descriptions.\\ 
Radiologists' eye gaze is a valuable modality that provides insights into radiologists' differential focus on image regions. Recent studies highlight the potential of leveraging radiologists' visual search patterns to enhance disease diagnosis~\cite{bhattacharya2022gazeradar, radiotransformer, bhattacharya2024gazediff}. 
These approaches provide evidence that radiologists’ visual search patterns can guide models in prioritizing relevant image regions; we hypothesize that this can lead to more accurate and contextually aligned report generation. 
Despite the recent progress in incorporation of gaze into ML models, research on integrating eye-tracking data for medical report generation remains limited, primarily due to challenges in multimodal integration complexity and the cost of acquiring eye gaze data during model deployment. EGGCA~\cite{eggca} is the sole notable work in this area, developing several modules for aligning eye gaze, image, and text at multiple levels. However, this method does not exploit the full potential of the modality interactions, leading to potentially inferior report generation capabilities.\\
We introduce \textit{Gaze2Report}, a novel architecture that integrates eye gaze into radiology report generation.
The main contributions of this study are threefold:\\
a) We employ GNN to enhance the visual-gaze modality interaction in a radiology report generation framework. By combining these visual-gaze tokens with generation instructions, we also improve the LLM’s ability to assess modality relevance, leading to more accurate report generation.\\
b) \textit{Gaze2Report} demonstrates that combining both image and gaze for report generation outperforms using the image alone. Additionally, to address the absence of gaze during inference, we employ an auxiliary scanpath prediction module, enhancing performance and providing better explainability.\\
c) The model's performance is evaluated on multiple datasets using both NLG and clinical efficacy (CE) metrics to validate its effectiveness.
\section {Methodology}
\noindent\textbf{Overview.} Our proposed \textit{Gaze2Report} framework consists of two main components: a visual-gaze token generation module and a LLM decoder for text generation. The visual-gaze module involves four steps: 1) extracting visual features from images using a Vision Transformer (ViT)~\cite{vit}, 2) time-aggregating gaze fixations from MedGaze~\cite{medgaze} to generate gaze embeddings, 3) passing visual, gaze, and positional features through a GNN to obtain fused visual-gaze embeddings, and 4) projecting graph embeddings into the high-dimensional LLM feature space via linear layers. Simultaneously, the LLM tokenizer processes ``generation instructions'' and reports. Finally, these embeddings are concatenated into a unified multimodal prompt to fine-tune the LoRA layers of the LLM for autoregressive report generation. During training, only the report tokens are included in the autoregressive process. Next, we discuss the main components of \textit{Gaze2Report} in detail (see Fig.~\ref{framework}).

\noindent \textbf{Visual token generation.} Given an input chest X-ray image \( I \in \mathbb{R}^{H \times W} \), we first patchify it into \( n \) non-overlapping grids, where each patch \( P_i \in \mathbb{R}^{h \times w} \) (for \( i = 1, 2, \dots, n \)) is a small section of the image. These patches are then passed through a ViT to extract spatial features, denoted as \( V_{I} \in \mathbb{R}^{n \times D} \), where \( D \) is the dimensionality of each patch's feature vector.\\
\textbf{Gaze token generation.} Next, we employ a scanpath prediction model, MedGaze, to simulate a sequence of gaze fixation points, capturing their spatial coordinates and temporal duration. This process enables the identification of specific regions within the image that attract the radiologist's visual attention, facilitating a deeper understanding of their interpretative focus. To capture this attention focused on each patch, we perform duration aggregation, where the fixation durations of all eye-gaze points within each patch are summed up. Specifically, for each patch \( P_i \), the fixation time \( T_i \) is computed by summing the fixation durations \( t_{ij} \) of all eye-gaze points \( j \) falling within that patch as $T_i = \sum_{j \in P_i} t_{ij}$. The aggregated fixation time \( T_i \) is then replicated into a gaze vector embedding \( G_i \), which reflects the radiologist’s attention distribution over the image patch.\\ 
\textbf{GNN for multimodal interaction.} A graph is constructed for the image such that each node represents a patch with associated visual, gaze tokens and positional encodings. Visual token \( V_i \) and gaze token \( G_i \) generation is explained in the previous sections. A learnable positional encoding \( e_i \) is included that indicates the patch's location in the image. The node feature vector for each patch \( i \) is then represented as \( h_i = \text{Concat}(V_i, G_i, p_i) \).\\
The dot product between the positional encodings of patches \(i\) and \(j\), $e_i^T e_j$, is used to identify the k-nearest neighbors (k-NN) for graph construction. Edges are formed between nodes based on their relative positional distance, as determined by the k-NN criterion. A GNN with multiple graph processing blocks~\cite{gnn2}, is then applied to update the node embeddings iteratively. Each block contains multiple fully connected (FC) layers along with a graph convolutional layer.
\[
h_{i}^{(t+1)} = \text{GNN}^{(t)} \left( h_{i}^{(t)}, \{ h_{j}^{(t)} : j \in N(i) \} \right)
\]
where \( N(i) \) denotes the neighbors of node \( i \) and \( h_i^{(t)} \) refers to the feature vector of node \( i \) at iteration \( t \). 
Once the node features have been updated, the entire graph's representation \( h_{\text{graph}} \) can be obtained by mean pooling the final node embeddings across the graph. $h_{\text{graph}} = \frac{1}{n} \sum_{i=1}^{n} h_i^{(T)}$, where \(n\) is the total number of nodes in the graph.

\noindent\textbf{Tokenization of Visual-gaze, Instruction, and Reports.} The instruction prompt
\( X_p \), such as ``Generate a comprehensive and detailed diagnosis report for this chest X-ray image,'' is tokenized using the Llama2 tokenizer into tokens: $
T_I = Tokenizer(X_p) = [t_{I1}, t_{I2}, ..., t_{Ik}]
$
where \( k \) is the number of tokens generated. The ground truth report \( X_r \) is similarly tokenized into: $
T_{R} = Tokenizer(X_r) = [t_{R1}, t_{R2}, ..., t_{2}, ..., t_{Rm}]
$
where \( m \) is the number of tokens in the report. To align the extracted graph features with the LLM's feature space, we apply a linear projection layer to project it into a high-dimensional space: $H_{\text{proj}} = Linear (h_{graph})$.\\
\textbf{Concatenation and visual-gaze prompt tuning of LLM.} The instruction tokens \( T_I \), report tokens \( T_R \), and visual-gaze tokens \( H_{\text{proj}} \) are concatenated to form a unified multimodal prompt for the Llama2-7B model: $X_{\text{input}} = [T_I, H_{\text{proj}}, T_R]$. For an image \(I\) with report \(X_r\)
, the detailed prompt input into Llama2 is:
$ \scriptstyle \texttt{[PROMPT]} \, Human: \langle Img:Gaze \rangle \, H_{proj} \, \langle /Img:Gaze \rangle, T_I.  \textbackslash n Assistant: T_R \, \texttt{[/PROMPT]}$.\\
Llama2-7B is employed for radiology report generation. Instead of updating all the parameters of the LLM, we focus on fine-tuning a smaller set of low-rank adapters, which is both computationally efficient and effective for adapting the model to the specific task of radiology report generation. The model processes the multimodal input sequence \( X_{input} \) and predicts the next report token based on prior tokens. The output sequence \( X_{output} \) is then compared with the ground truth report tokens \( T_R \) during training to optimize the model's performance. We use an autoregressive loss function that is designed to minimize the negative log-likelihood of predicting the next token in the sequence.
$\mathcal{L}(\theta; X_r, X_v, X_p) = -\sum_{i=1}^L \log p_\theta(x_i \mid X_v, X_p, X_r, x_{<i})
$
where \(\theta\) represents the trainable parameters of the model, \( L \) is the report length, \( x_i \) is the current prediction token, and \( x_{<i} \) represents all previous tokens.

\begin{figure*}[t]
\centering
\includegraphics[width=.7\textwidth]{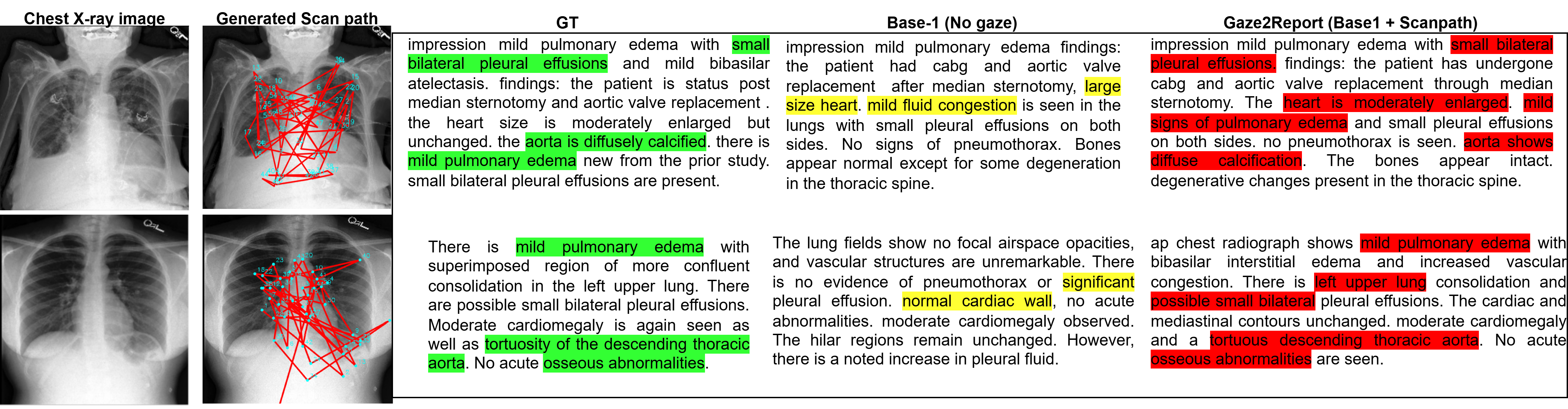}
\caption{Examples of the generated report on the MIMIC-CXR dataset, compared against Base-1, which does not utilize gaze. Our model successfully generated critical terms such as \textit{small bilateral pleural effusions}, \textit{aorta shows diffuse calcification}, \textit{left upper lung}, and \textit{osseous abnormalities}, which were absent in the baseline-generated report. 
Additionally, our method demonstrates greater clinical precision, using terms like \textit{mild signs of pulmonary edema} instead of \textit{mild fluid congestion}.
} \label{vis}
\end{figure*}
\begin{table*}[t]
\centering
\scriptsize
\caption{Comparison with SOTA in terms of NLG metrics. B, RG-L, and MTR represent BLEU, ROUGE, and METEOR metrics, respectively. Best and second best scores are \textbf{bolded} and \underline{underlined}, respectively}
\label{tab:main}

\resizebox{0.7\textwidth}{!}{
\begin{tabular}{c|cccccc|cccccc}
\hline
\cellcolor[HTML]{FFFFFF}{\color[HTML]{000000} }                                   & \multicolumn{6}{c|}{\cellcolor[HTML]{FFFFFF}{\color[HTML]{000000} \textbf{IU-XRAY}}}                                                                                                                                                                                                                            & \multicolumn{6}{c}{{\color[HTML]{000000} \textbf{MIMIC-CXR}}}                                                                                                                                                                                                                                                   \\ \cline{2-13} 
\multirow{-2}{*}{\cellcolor[HTML]{FFFFFF}{\color[HTML]{000000} \textbf{Methods}}} & \cellcolor[HTML]{FFFFFF}{\color[HTML]{000000} \textbf{B-1}} & \cellcolor[HTML]{FFFFFF}{\color[HTML]{000000} \textbf{B-2}} & \cellcolor[HTML]{FFFFFF}{\color[HTML]{000000} \textbf{B-3}} & {\color[HTML]{000000} \textbf{B-4}}   & {\color[HTML]{000000} \textbf{RG-L}}  & {\color[HTML]{000000} \textbf{MTR}}   & \cellcolor[HTML]{FFFFFF}{\color[HTML]{000000} \textbf{B-1}} & \cellcolor[HTML]{FFFFFF}{\color[HTML]{000000} \textbf{B-2}} & \cellcolor[HTML]{FFFFFF}{\color[HTML]{000000} \textbf{B-3}} & {\color[HTML]{000000} \textbf{B-4}}   & {\color[HTML]{000000} \textbf{RG-L}}  & {\color[HTML]{000000} \textbf{MTR}}   \\ \hline
\cellcolor[HTML]{FFFFFF}{\color[HTML]{000000} \textbf{ST \cite{showtell}}}                & \cellcolor[HTML]{FFFFFF}{\color[HTML]{000000} 0.243}        & \cellcolor[HTML]{FFFFFF}{\color[HTML]{000000} 0.130}        & \cellcolor[HTML]{FFFFFF}{\color[HTML]{000000} 0.108}        & {\color[HTML]{000000} 0.078}          & {\color[HTML]{000000} 0.307}          & {\color[HTML]{000000} 0.157}          & {\color[HTML]{000000} 0.308}                                & {\color[HTML]{000000} 0.190}                                & {\color[HTML]{000000} 0.125}                                & {\color[HTML]{000000} 0.088}          & {\color[HTML]{000000} 0.256}          & {\color[HTML]{000000} 0.122}          \\
\cellcolor[HTML]{FFFFFF}{\color[HTML]{000000} \textbf{Att2in \cite{Att2in}}}           & \cellcolor[HTML]{FFFFFF}{\color[HTML]{000000} 0.248}        & \cellcolor[HTML]{FFFFFF}{\color[HTML]{000000} 0.134}        & \cellcolor[HTML]{FFFFFF}{\color[HTML]{000000} 0.116}        & {\color[HTML]{000000} 0.091}          & {\color[HTML]{000000} 0.309}          & {\color[HTML]{000000} 0.162}          & {\color[HTML]{000000} 0.314}                                & {\color[HTML]{000000} 0.198}                                & {\color[HTML]{000000} 0.133}                                & {\color[HTML]{000000} 0.095}          & {\color[HTML]{000000} 0.264}          & {\color[HTML]{000000} 0.122}          \\
\cellcolor[HTML]{FFFFFF}{\color[HTML]{000000} \textbf{AdaAtt \cite{AdaAtt}}}            & \cellcolor[HTML]{FFFFFF}{\color[HTML]{000000} 0.284}        & \cellcolor[HTML]{FFFFFF}{\color[HTML]{000000} 0.207}        & \cellcolor[HTML]{FFFFFF}{\color[HTML]{000000} 0.150}        & {\color[HTML]{000000} 0.126}          & {\color[HTML]{000000} 0.311}          & {\color[HTML]{000000} 0.165}          & {\color[HTML]{000000} 0.314}                                & {\color[HTML]{000000} 0.198}                                & {\color[HTML]{000000} 0.132}                                & {\color[HTML]{000000} 0.094}          & {\color[HTML]{000000} 0.267}          & {\color[HTML]{000000} 0.128}          \\
{\color[HTML]{000000} \textbf{R2Gen \cite{r2gen}}}                                   & {\color[HTML]{000000} 0.470}                                & {\color[HTML]{000000} 0.304}                                & {\color[HTML]{000000} 0.219}                                & {\color[HTML]{000000} 0.165}          & {\color[HTML]{000000} 0.371}          & {\color[HTML]{000000} 0.187}          & {\color[HTML]{000000} 0.353}                                & {\color[HTML]{000000} 0.218}                                & {\color[HTML]{000000} 0.145}                                & {\color[HTML]{000000} 0.103}          & {\color[HTML]{000000} 0.277}          & {\color[HTML]{000000} 0.142}          \\
{\color[HTML]{000000} \textbf{XPRONET \cite{xpronet}}}                                  & {\color[HTML]{000000} 0.478}                                & {\color[HTML]{000000} 0.308}                                & {\color[HTML]{000000} 0.223}                                & {\color[HTML]{000000} 0.168}          & {\color[HTML]{000000} 0.376}          & {\color[HTML]{000000} 0.190}          & {\color[HTML]{000000} 0.346}                                & {\color[HTML]{000000} 0.213}                                & {\color[HTML]{000000} 0.147}                                & {\color[HTML]{000000} 0.104}          & {\color[HTML]{000000} 0.280}          & {\color[HTML]{000000} 0.136}          \\
{\color[HTML]{000000} \textbf{EGGCA-Net \cite{eggca}}}                             & {\color[HTML]{000000} 0.482}                                & {\color[HTML]{000000} 0.314}                                & {\color[HTML]{000000} 0.224}                                & {\color[HTML]{000000} 0.167}          & {\color[HTML]{000000} 0.372}          & {\color[HTML]{000000} 0.186}          & {\color[HTML]{000000} 0.364}                                & {\color[HTML]{000000} 0.228}                                & {\color[HTML]{000000} 0.157}                                & {\color[HTML]{000000} 0.107}          & {\color[HTML]{000000} 0.278}          & {\color[HTML]{000000} 0.146}          \\
{\color[HTML]{000000} \textbf{MET\cite{met}}}                                       & {\color[HTML]{000000} 0.483}                                & {\color[HTML]{000000} \underline{0.322}}                       & {\color[HTML]{000000} 0.228}                                & {\color[HTML]{000000} 0.172}          & {\color[HTML]{000000} \underline{0.380}} & {\color[HTML]{000000} 0.192}          & {\color[HTML]{000000} 0.386}                                & {\color[HTML]{000000} 0.250}                                & {\color[HTML]{000000} 0.169}                                & {\color[HTML]{000000} 0.124}          & {\color[HTML]{000000} 0.291}          & {\color[HTML]{000000} 0.152}          \\
{\color[HTML]{000000} \textbf{R2GenGPT \cite{r2gengpt}}}                                & {\color[HTML]{000000} \underline{0.488}}                       & {\color[HTML]{000000} 0.316}                                & {\color[HTML]{000000} \underline{0.228}}                       & {\color[HTML]{000000} \underline{0.173}} & {\color[HTML]{000000} 0.377}          & {\color[HTML]{000000} \underline{0.211}} & {\color[HTML]{000000} \underline{0.411}}                       & {\color[HTML]{000000} \underline{0.267}}                       & {\color[HTML]{000000} \underline{0.186}}                       & {\color[HTML]{000000} \underline{0.134}} & {\color[HTML]{000000} \underline{0.297}} & {\color[HTML]{000000} \underline{0.160}} \\ \hline
{\color[HTML]{000000} \textbf{Base-1}}                              & {\color[HTML]{000000} 0.484}                                & {\color[HTML]{000000} 0.318}                                & {\color[HTML]{000000} 0.226}                                & {\color[HTML]{000000} 0.170}          & {\color[HTML]{000000} 0.376}          & {\color[HTML]{000000} 0.206}          & {\color[HTML]{000000} 0.407}                                & {\color[HTML]{000000} 0.262}                                & {\color[HTML]{000000} 0.183}                                & {\color[HTML]{000000} 0.132}          & {\color[HTML]{000000} 0.294}          & {\color[HTML]{000000} 0.157}          \\
{\color[HTML]{000000} \textbf{Base-2}}                                   & {\color[HTML]{000000} 0.491}                                & {\color[HTML]{000000} 0.323}                                & {\color[HTML]{000000} 0.231}                                & {\color[HTML]{000000} 0.175}          & {\color[HTML]{000000} 0.381}          & {\color[HTML]{000000} 0.213}          & {\color[HTML]{000000} 0.413}                                & {\color[HTML]{000000} 0.268}                                & {\color[HTML]{000000} 0.188}                                & {\color[HTML]{000000} 0.135}          & {\color[HTML]{000000} 0.298}          & {\color[HTML]{000000} 0.163}          \\
\textbf{Base-3}                                                      & {\color[HTML]{000000} 0.493}                                & {\color[HTML]{000000} 0.323}                                & {\color[HTML]{000000} 0.232}                                & {\color[HTML]{000000} 0.176}          & {\color[HTML]{000000} 0.383}          & {\color[HTML]{000000} 0.215}          & {\color[HTML]{000000} 0.414}                                & {\color[HTML]{000000} 0.269}                                & {\color[HTML]{000000} 0.190}                                & {\color[HTML]{000000} 0.136}          & {\color[HTML]{000000} 0.301}          & {\color[HTML]{000000} 0.165}          \\ \hline
\textbf{\textit{Gaze2Report}}                                                                     & {\color[HTML]{000000} \textbf{0.495}}                       & {\color[HTML]{000000} \textbf{0.326}}                       & {\color[HTML]{000000} \textbf{0.235}}                       & {\color[HTML]{000000} \textbf{0.178}} & {\color[HTML]{000000} \textbf{0.386}} & {\color[HTML]{000000} \textbf{0.218}} & {\color[HTML]{000000} \textbf{0.419}}                       & {\color[HTML]{000000} \textbf{0.272}}                       & {\color[HTML]{000000} \textbf{0.192}}                       & {\color[HTML]{000000} \textbf{0.137}} & {\color[HTML]{000000} \textbf{0.303}} & {\color[HTML]{000000} \textbf{0.167}} \\ \hline

\end{tabular}
}
\end{table*}
\section{Experiment Design and Results}
\begin{table}[t]
\centering
\begin{subtable}[t]{0.4\textwidth}
\centering
\captionsetup{width=\textwidth}
\caption{Report generation on the REFLACX dataset.}
\label{tab:reflacx}
\scriptsize
\resizebox{0.8\textwidth}{!}{
\begin{tabular}{c|c|ccccc}
\hline
 &  & \multicolumn{5}{c}{\textbf{REFLACX}} \\ \cline{3-7} 
\multirow{-2}{*}{\textbf{\begin{tabular}[c]{@{}c@{}}Gaze\\ (Inf.)\end{tabular}}} 
 & \multirow{-2}{*}{\textbf{Methods}} 
 & \textbf{B-1} & \textbf{B-2} & \textbf{B-3} & \textbf{B-4} & \textbf{MTR} \\ \hline

\textbf{Absent} & \textbf{Base-1} 
& \color[HTML]{6434FC}{\textbf{0.354}}
& \color[HTML]{6434FC}{\textbf{0.226}}
& \color[HTML]{6434FC}{\textbf{0.163}}
& \color[HTML]{6434FC}{\textbf{0.109}}
& \color[HTML]{6434FC}{\textbf{0.151}} \\ \hline

 & \textbf{Mask}   & 0.365 & 0.234 & 0.168 & 0.115 & 0.158 \\
 & \textbf{Unmask} & 0.366 & 0.237 & 0.170 & 0.117 & 0.160 \\
 & \textbf{Base-2} & 0.372 & 0.240 & 0.174 & 0.120 & 0.164 \\
 & \textbf{Base-3} & 0.374 & 0.242 & 0.176 & 0.122 & 0.165 \\
\multirow{-5}{*}{\textbf{Present}} 
 & \textbf{\textit{Gaze2Report}}
 & \color[HTML]{32CB00}{\textbf{0.375}}
 & \color[HTML]{32CB00}{\textbf{0.243}}
 & \color[HTML]{32CB00}{\textbf{0.177}}
 & \color[HTML]{32CB00}{\textbf{0.123}}
 & \color[HTML]{32CB00}{\textbf{0.167}} \\ \hline

 & \textbf{Base-2} & 0.363 & 0.233 & 0.168 & 0.114 & 0.157 \\
 & \textbf{Base-3} & 0.367 & 0.236 & 0.168 & 0.116 & 0.159 \\
\multirow{-3}{*}{\textbf{Absent}}
 & \textbf{\textit{Gaze2Report}}
 & \color[HTML]{FF0000}{0.369}
 & \color[HTML]{FF0000}{0.237}
 & \color[HTML]{FF0000}{0.170}
 & \color[HTML]{FF0000}{0.117}
 & \color[HTML]{FF0000}{0.161} \\ \hline
\end{tabular}
}
\end{subtable}
\hfill
\begin{subtable}[t]{0.3\textwidth}
\centering
\scriptsize
\caption{Report generation on the MIMIC-CXR dataset.}
\label{tab:clinical}
\resizebox{\textwidth}{!}{
\begin{tabular}{c|cccc}
\hline
 & \multicolumn{4}{c}{\textbf{MIMIC-CXR}} \\ \cline{2-5}
\textbf{Methods} 
 & \textbf{Acc} & \textbf{Pre} & \textbf{Re} & \textbf{F1} \\ \hline

\textbf{R2Gen}     & 0.728 & 0.396 & 0.307 & 0.346 \\
\textbf{EGGCA-Net} & 0.706 & \textbf{0.569} & \textbf{0.535} & \textbf{0.551} \\
\textbf{MET}       & 0.753 & 0.364 & 0.309 & 0.334 \\
\textbf{R2GenGPT}  & 0.772 & 0.392 & 0.387 & 0.389 \\ \hline

\textbf{Base-1} & 0.759 & 0.405 & 0.384 & 0.393 \\
\textbf{Base-2} & 0.773 & 0.442 & 0.417 & 0.428 \\
\textbf{Base-3} & \underline{0.781} & 0.450 & 0.426 & 0.437 \\ \hline

\textbf{\textit{Gaze2Report}} 
 & \textbf{0.786} & \underline{0.457} & \underline{0.431} & \underline{0.444} \\ \hline
\end{tabular}
}
\end{subtable}

\caption{Comparison of report generation across REFLACX (left) and MIMIC-CXR (right).}
\end{table}

\textbf{Dataset description.} This study utilizes three datasets: REFLACX~\cite{reflacx}, IU-XRAY~\cite{iuxray}, and MIMIC-CXR~\cite{mimic}.  REFLACX includes CXRs with synchronized eye-tracking and transcription pairs annotated by radiologists (1800 train, 707 test samples). IU-XRAY contains 3955 fully de-identified radiology reports associated with 7470 CXRs, including both frontal and lateral views. Following~\cite{partition}, we use a 7:1:2 train/test/val split with evaluations conducted on the test split. MIMIC-CXR consists of 377,110 CXRs and 227,835 reports from 64,588 patients. We conformed to the standard partitioning~\cite{partition} of using 270,790 samples for training, 2,130 for validation, and 3,858 for testing.\\ 
\textbf{Metrics.} To gauge the linguistic quality and factual consistency of generated reports, we use natural language generation metrics (NLG) and clinical efficacy metrics (CE). For NLG, we use BLEU~\cite{bleu}, METEOR~\cite{meteor}, and ROUGE-L~\cite{rouge}. We also report CE to assess the accuracy of descriptions for clinical abnormalities. Regarding this, we label reports with Chexbert~\cite{chexbert}, assigning binary labels for fourteen thoracic pathologies, and compute Accuracy, Precision, Recall, and F1 Score for each disease class.\\
\textbf{Implementation Details.} \textit{Gaze2Report} is implemented in PyTorch using a 48 GB Nvidia Quadro RTX 8000 GPU. Images are resized to 224 $\times$ 224, and non-overlapping patches of size 16 $\times$ 16 are extracted. The Llama2-7B model is used as the LLM, with ImageNet-21k pretrained ViTB-16 as the visual encoder. For LoRA, the attention dimension and scaling parameter are both set to 16. Experiments are conducted for 5 epochs on the MIMIC-CXR dataset and 18 epochs on the IU-Xray dataset, with a mini-batch size of 8. During testing, a beam search strategy with a beam size of 3 is applied. The Adam optimizer is used with a linear decay scheduler and an initial learning rate of \(5 \times 10^{-4}\).\\
\textbf{NLG Evaluation.} In Table~\ref{tab:main}, we compare the proposed \textit{Gaze2Report} framework with standard image captioning methods (ST~\cite{showtell}, Att2in~\cite{Att2in}, AdaAtt~\cite{AdaAtt}) and medical report generation models (XPRONET~\cite{xpronet}, EGGCA-Net~\cite{eggca}, METransformer~\cite{met}, R2Gen~\cite{r2gen}, and R2GenGPT~\cite{r2gengpt}). Additionally, we introduce three baselines: 1) \textbf{Base-1}: Relies only on visual embeddings, without gaze embeddings generated via the scanpath prediction module. 2) \textbf{Base-2}: Concatenates visual and gaze embeddings before input to the LLM. 3) \textbf{Base-3}: Passes visual and gaze embeddings through a cross-attentional fusion module to create visual-gaze tokens. On the MIMIC-CXR dataset, \textit{Gaze2Report} outperforms the closest competitors, MET and R2GenGPT, across all metrics. MET uses multiple learnable expert tokens within a transformer-based framework, while R2GenGPT processes visual and instruction tokens into the LLM, similar to Base-1. Both methods fail to integrate medical domain knowledge, resulting in lower performance. Though EGGCA-Net utilizes gaze heatmaps via multiple alignment modules, it does not achieve the same level of interaction as GNN, which incorporates fixation duration and coordinates. Moreover, EGGCA-Net lacks the generalization capability of the pretrained Llama2-7B model. BLEU-n (n = 1, 2, 3, 4) scores improved by 0.08, 0.05, 0.06, and 0.03 over the second-best approaches. Likewise, ROUGE and METEOR scores saw improvements of 0.06 and 0.07, respectively. Similar trends were observed on the IU-XRAY dataset. \textit{Gaze2Report} also outperformed Base-1, which only uses visual embeddings, confirming that incorporating scanpaths can lead to more accurate report generation.
Additionally, the improvement over baseline fusion models (Base-2 and Base-3) in BLEU-1 (+0.05), BLEU-4 (+0.01), METEOR (+0.02), and ROUGE (+0.02) scores further supports the importance of the GNN-based visual-gaze assembly module in our architecture.\\
\textbf{CE Evaluation.} Table~\ref{tab:clinical} compares \textit{Gaze2Report} with radiology report generation models (R2Gen, R2GenGPT, EGGCA-Net, MET) based on CE scores. \textit{Gaze2Report} surpasses R2GenGPT in Acc (+0.014), Prec (+0.065), Rec (+0.044), and F1 Score (+0.055), demonstrating its ability to generate crucial clinical information. Only EGGCA-Net outperforms our model due to its dedicated multi-label classification losses for abnormalities (e.g., pneumonia, pleural effusion, cardiomegaly). Also, \textit{Gaze2Report} achieves a higher BERTScore (0.241) and RadGraph F1 score~\cite{radgraph} (0.117) compared to R2Gen (0.183, 0.062), with only EGGCA-Net achieving superior results (0.379, 0.148). High BERTScore and RadGraph F1 scores indicate strong semantic similarity, coherence, and precise representation of clinical terminologies in our generated reports. Qualitative results in Fig.~\ref{vis}.\\
\textbf{Corollary Experiment.}
Table~\ref{tab:reflacx} presents an experiment demonstrating that \textit{Gaze2Report} and its variants (Base-2, Base-3) outperform Base-1 when gaze is absent at inference, aided by the scanpath prediction module (0.117 vs. 0.109 BLEU-4, 0.161 vs. 0.151 METEOR). Additionally, \textit{Gaze2Report} outperforms the Mask and Unmask baselines, which use gaze-based attention (manual or learnable) to filter out irrelevant visual input patches. This holds true even though these baselines have access to gaze during inference, while \textit{Gaze2Report} does not. We also demonstrate the generalizability of \textit{Gaze2Report} - in the presence of gaze at inference, it can directly extract gaze embeddings from input eye gaze data without relying on scanpath prediction, and still outperforms both baselines.

\section{Conclusion}
Current methods often struggle to generate radiology reports with strong alignment between impressions and disease manifestations due to the absence of physician-informed medical priors such as eye gaze. To address this, we enhance feature relevance through rich visual-gaze interaction via GNN and improve interpretability by tuning an LLM with a multimodal visual-gaze-instruction prompt. Additionally, we mitigate the challenge of gaze being absent at inference by incorporating an auxiliary scanpath prediction task. 

\section{Compliance with Ethical Standards}
\vspace{-1mm}
This study, conducted on open-source data, did not require ethical approval.\\
\textbf{Acknowledgements:}
This research was partially supported by NIH grants 1R01CA297843-01, 1R03DE033489-01A1, and NSF grant 2442053. The content is solely the responsibility of the authors and does not necessarily represent the official views of the NIH.

\bibliographystyle{IEEEbib}
\bibliography{strings,refs}

\end{document}